\documentclass{ws-ijmpa}
\usepackage{graphicx}

\begin{document}

\markboth{A.V.Minkevich} {De Sitter spacetime with torsion as
physical spacetime in the vacuum and isotropic cosmology}

\catchline{}{}{}{}{}
%

\title{De Sitter spacetime with torsion as
physical spacetime in the vacuum and isotropic cosmology}

\author{A.V. Minkevich}
 \address{Department of Theoretical Physics, Belarussian State University,\\
 Minsk, 220030, Belarus\\
minkav@bsu.by}
  \address{Department of Physics and Computer Methods, Warmia and Mazury University in
 Olsztyn,\\
 Olzstyn, Poland\\
 awm@matman.uwm.edu.pl}

\maketitle

\begin{history}
\received{18 October 2010}
\end{history}

\begin{abstract}
Homogeneous isotropic models with two torsion functions built in
the framework of the Poincar\'e gauge theory of gravity (PGTG)
based on general expression of gravitational Lagrangian without
cosmological constant are analyzed. It is shown that the physical
spacetime in the vacuum in the frame of PGTG can have the
structure of flat de Sitter spacetime with torsion. Some physical
consequences of obtained conclusion are discussed.
\end{abstract}

\ccode{PACS numbers: 04.50.+h; 98.80.Cq; 11.15.-q; 95.36.+x}

\section{Introduction}

Since the creation of the special relativity theory the physical spacetime in
the vacuum (without physical fields) is considered as Minkowski spacetime with
the structure of pseudo-Euclidien continuum. According to the general
relativity theory (GR), the physical spacetime in the gravitational field
possesses the structure of pseudo-Riemannien continuum, however, far from
gravitating objects and in absence of gravitational waves the physical
spacetime in GR in fact can be considered as Minkowski spacetime \footnote{Here
and later we assume that cosmological constant is equal to zero.}. In the
framework of the Poincar\'e gauge theory of gravity \cite{a1,a2,a3,a4,a5,a6},
which is a natural and in certain sense necessary generalization of Einsteinian
GR (see Refs. 7, 8 and references herein), the physical spacetime in the
gravitational field has the structure of Riemann-Cartan continuum. Usually one
supposes that in the frame of PGTG, similarly to GR, far from gravitating
objects and in absence of gravitational waves the properties of physical
spacetime are practically the same that of Minkowski spacetime. However, as it
will be shown in this paper, the situation in PGTG can be essentially
different, and the structure of physical spacetime in the vacuum can vary from
that of Minkowski spacetime which leads to important physical consequences.

 Before we analyze the corresponding situation in the
framework of PGTG, it is necessary to note why this is of direct
physical interest for the gravitation theory, modern cosmology and
astrophysics. As it was shown in a number of papers (see
Refs.~\refcite{a9}--\refcite{a15} and references herein), the PGTG
offers opportunities to solve principal cosmological problems --
the problem of cosmological singularity, the problem of dark
components of the Universe -- dark energy and dark matter by
describing the gravitational field in 4-dimensional classical
physical space-time. It is because the PGTG leads to essential
changes of gravitational interaction in comparison with GR and
Newton's theory of gravity by certain physical conditions, in
particular at extreme conditions (extremely high energy densities
and pressures) in the beginning of cosmological expansion. These
changes are connected with the structure of physical spacetime in
PGTG, namely with spacetime torsion, the presence of which is a
necessary consequence of including the Lorentz group to the gauge
group which corresponds to gravitational interaction.

The present paper is organized in the following way. In Section~\ref{secii}, we
present the main relations for homogeneous isotropic models (HIM) obtained in
the framework of PGTG based on general expression of gravitational Lagrangian.
In Section~\ref{seciii}, the structure of spacetime in the vacuum in PGTG is
analyzed. In Conclusion, some physical consequences connected with the
structure of physical spacetime in the vacuum are discussed.

\section{Homogeneous Isotropic Models in PGTG \label{secii}}

From the physical point of view the spacetime in the vacuum is homogeneous and
isotropic, and in order to investigate the structure of vacuum spacetime in the
frame of PGTG we will analyze the HIM. We will consider the PGTG based on
general expression of gravitational Lagrangian ${\cal L}_{\rm g}$ including
both a scalar curvature and various invariants quadratic in gravitational gauge
field strengths -- the curvature and torsion tensors (definitions and notations
of Ref. 14 are used below):
\begin{eqnarray} \label{1}
 {\cal L}_{\rm g}=f_0\,
F+F^{\alpha\beta\mu\nu}(f_1\:F_{\alpha\beta\mu\nu}
                +f_2\:
                F_{\alpha\mu\beta\nu}+f_3\:F_{\mu\nu\alpha\beta})
        + F^{\mu\nu}( f_4\:F_{\mu\nu} \nonumber
 \\ + f_5\: F_{\nu\mu}) +  f_6\:F^2
        +S^{\alpha\mu\nu} (a_1\:S_{\alpha\mu\nu}+a_2\:
        S_{\nu\mu\alpha})
    +a_3\:S^\alpha{}_{\mu\alpha}S_\beta{}^{\mu\beta}.
\end{eqnarray}

The Lagrangian (1) includes the parameter $f_0=(16\pi G)^{-1}$ ($G$ is Newton's
gravitational constant, the light velocity $c=1$) and a number of indefinite
parameters: $f_i$ ($i=1,2,...6$) and $a_k$ ($k=1,2,3$). By using the expression
(1) for ${\cal L}_{\rm g}$ the system of gravitational equations for HIM filled
by gravitating matter with energy density $\rho$ and pressure $p$ was obtained
in Ref. 12 (see also Ref. 14). This system contains 4 differential equations
for three geometric characteristics of HIM as functions of time -- the scale
factor of Robertson-Walker metrics $R$ and two torsion functions $S_1$ and
$S_2$ \footnote{For the first time equations for HIM with two torsion functions
were deduced in Ref. 16. These equations were considered in \cite{a17} with the
purpose to obtain their solutions; however, so called "modified double duality
ansatz" used in Ref. 17 by obtaining solutions with nonvanishing torsion
function $S_2$ is not applicable in this case even for the vacuum (see below)
and its application in general leads to incorrect solutions.}. Generally the
system of gravitational equations for HIM contains 5 following indefinite
parameters:
\begin{eqnarray}
  a = 2a_1  + a_2  + 3a_3, \qquad b = a_2  - a_1, \qquad
  f = f_1  + \frac{{f_2 }} {2} + f_3  + f_4  + f_5  + 3f_{6}\, ,
\nonumber\\
  q_1  = f_2  - 2f_3  + f_4  + f_5  + 6f_{6}, \qquad q_2  = 2f_1  - f_2 .
\nonumber
\end{eqnarray}
Gravitational equations for HIM allow to obtain cosmological
equations generalizing Friedmann cosmological equations of GR in
the following form:
\begin{eqnarray}\label{2}
    \dot{H}+H^2-2HS_1-2\dot{S}_1 = A_1,
\end{eqnarray}
\begin{eqnarray}\label{3}
    \frac{k}{R^2} + (H-2S_1)^2 - S_2^2 = A_2,
\end{eqnarray}
where the curvature functions $A_1$ and $A_2$ are determined from
gravitational equations by the following way:
\begin{eqnarray}\label{4}
A_1=-\frac{1} {{12(f_0 +a/8) Z}}
        \Big[
            \rho  + 3p - \frac{2f}{3} F^2 + 8 q_2 FS_2^2
\nonumber\\
      - 12q_2 \left( {\left( {HS_2  + \dot S_2 } \right)^2
                + 4\left( {\frac{k}{{R^2 }} - S_2^2 } \right)S_2^2 }
            \right) - \frac{3a} {2}
            \left( \dot{H} + H^2 \right)
            \Big],
\nonumber\\
A_2=\frac{1} {{6(f_0 +a/8)Z}}
       \Big[
            \rho  - 6 (b +a/8)S_2^2 + \frac{f}{3} F^2
        + \frac{3a} {4}
            \left({\frac{k}{{R^2 }} + H^2}\right) \nonumber\\
             - 6 q_2 \left({\left( {HS_2  + \dot S_2 } \right)^2
                + 4\left( {\frac{k}{{R^2 }} - S_2^2 } \right)S_2^2
                }\right)
           \Big],
\end{eqnarray}
\noindent $H=\dot{R}/R $ is the Hubble parameter (a dot denotes
the differentiation with respect to time), the scalar curvature
$F=6(A_1+A_2)$ is
\begin{eqnarray}\label{5}
         F=\frac{1}{2(f_0 + a/8)} \Big[
                \rho-3p - 12(b+a/8) S_2^2
                + \frac{3a}{2} \left(\frac{k}{R^2}+\dot{H}+2H^2\right)
            \Big],
\end{eqnarray}
and $Z=1 + \frac{1} {(f_0+ a/8)} \left(\frac{2f} {3} F - 4q_2
S_2^2\right)$. Cosmological equations (2)-(3) contain the torsion
functions $S_1$ and $S_2$ with their first derivatives. From
gravitational equations for HIM the torsion function $S_1$ is
determined as
\begin{eqnarray} \label{6}
    S_1  = -\frac{1}{6 (f_0 + a/8)Z}
    \left[ f \dot F + 6(2f-q_1+2q_2) H S_2^2
            +6(2f-q_1) S_2 \dot S_2 \right],
\end{eqnarray}
and the torsion function $S_2$ satisfies the differential equation
of the second order:
\begin{eqnarray}\label{7}
    q_2 \left[ \ddot S_2  + 3H\dot S_2  + \left(3\dot{H} - 4 \dot S_1
    +4S_1(3H
        - 4 S_1)\right) S_2  \right]
\nonumber \\
        -  \left[\frac{q_1+q_2} {3} F + (f_0-b)
        -2 (q_1+q_2-2f) A_2 \right]S_2  = 0.
\end{eqnarray}
By given equation of state for gravitating matter cosmological
equations (2)-(3) together with equations (6)-(7) for torsion
functions describe the evolution of the most general HIM in PGTG.
So far, we have not used any restrictions on indefinite parameters
of ${\cal L}_{\rm g}$. From formulas (5)-(6) for scalar curvature
$F$ and torsion function $S_1$ we see that cosmological equations
(2)-(3) do not contain higher derivatives of the scale factor $R$
only if $a=0$. Isotropic cosmology with $a \neq 0$ possesses some
principal problems; in particular cosmological equations at
physically available initial conditions lead in this case to not
physical solutions\cite{a24}. With the purpose to exclude higher
derivatives of $R$ from cosmological equations the restriction
$a=0$ is used in our works. Because of mathematical reasons this
restriction will be not used by further general mathematical
analysis.

\section{Spacetime of gravitating vacuum in PGTG \label{seciii}}

Apart the spacial homogeneity and isotropy, in order to
investigate the gravitating vacuum in PGTG we have to take into
account also its homogeneity in time. In the frame of GR these
conditions in accordance with Friedmann cosmological equations (in
absence of gravitating  matter ($\rho=0$) and vanishing
cosmological constant) are fulfilled only in the case of flat HIM
($k=0$) with zeroth curvature that leads to Minkowski spacetime
\footnote{Such model filled with the dust ($p=0$) was considered
in \cite{a18} as preferable model of the real universe.}. However,
in the frame of PGTG there is also another solution, which can be
obtained if we suppose that in relations (2)-(7) for HIM the time
derivatives vanish and $\rho=0$. Then from (5) and (6) we obtain
the following expressions for scalar curvature $F$ and torsion
function $S_1$ in the vacuum ($k=0$):
\begin{eqnarray}\label{8}
         F=\frac{6}{f_0 + a/8} \left[
                 - (b+a/8) S_2^2
                + \frac{a}{4} H^2 \right],
                \nonumber \\
      S_1  = -\frac{2f - q_1 + 2q_2}{(f_0 + a/8)Z} H S_2^2,
\end{eqnarray}
and the curvature functions (4) in the vacuum take the following
form:
\begin{eqnarray}\label{9}
A_1= \frac{1} {{6(f_0 +a/8) Z}}
        \left[
             \frac{f}{3} F^2 - 4 q_2 FS_2^2
         +6q_2 \left( H^2 - 4 S_2^2 \right)S_2^2
             + \frac{3a} {4}
            H^2
            \right],
\nonumber\\
A_2=\frac{1} {6(f_0 +a/8)Z}
        \left[
             - 6 (b+a/8) S_2^2 + \frac{f}{3} F^2
        + \frac{3a} {4}H^2
             - 6 q_2 \left(H^2 - 4S_2^2 \right)S_2^2 \right].
\end{eqnarray}
Then cosmological equations (2)-(3) in the vacuum take the
following form:
\begin{eqnarray}\label{10}
H^2 \big[1 + \frac{2(2f - q_1 + 2q_2)}{(f_0 + a/8)Z} S_2^2\big] =
    A_1,
    \nonumber \\
    H^2 \big[1 + \frac{2(2f - q_1 + 2q_2)}{(f_0 + a/8)Z} S_2^2\big]^2 - S_2^2 =
    A_2.
\end{eqnarray}
Because the Bianchi identity for HIM \cite{a12} leads in this
particular case to the following relation:
$H(A_2-A_1)+2S_1A_1+HS_2^2=0$, only one out of eqs. (10) is
independent. Eq. (7) for $S_2$-function in the vacuum is
transformed to:
\begin{eqnarray}\label{11}
     S_2 [ 4q_2 S_1 (3H - 4S_1)
        -  \frac{q_1+q_2} {3} F
        +2 (q_1+q_2-2f) A_2 -(f_0-b)]  = 0,
\end{eqnarray}
where the curvature functions and $S_1$-function are determined by (8)-(9).
Eqs. (10)-(11) determine the values of $H$ and $S_2$ for gravitating vacuum. In
accordance with (11) there are two types of solutions: with vanishing and
nonvanishing value of $S_2$. If $S_2=0$, then we have $S_1=0$ and according to
cosmological equations in this case $H=0$. Such solution corresponds to
Minkowski spacetime in the vacuum. For the second solution with nonvanishing
value of $S_2$ we have:
\begin{eqnarray}\label{12}
      4q_2 S_1 (3H - 4S_1)
        -  \frac{q_1+q_2} {3} F
        +2 (q_1+q_2-2f) A_2 -(f_0-b)  = 0.
\end{eqnarray}
Eqs. (10) and (12) allow to determine the values of $S_2^2$ and $H^2$ for the
vacuum as functions of available indefinite parameters in gravitational
equations for HIM. This solution corresponds to 4-dimensional spatially flat de
Sitter spacetime with nonvanishing torsion i.e. to the Riemann-Cartan continuum
with constant curvature and torsion. It should be noted that the torsion in
discussed vacuum solutions is connected with pseudoscalar torsion function
$S_2$, and these solutions differ essentially from that obtained in the case of
HIM with the only torsion function $S_1$ \cite{a19}. According to Ref. 19 in
this case the vacuum solution with de Sitter metrics and nonvanishing torsion
is possible only if $a \neq 0$ \footnote{Such solution coincides with that
obtained in Ref. 20 in the case $f_0=0$.}. As it is follows from our
consideration, the regular character of the vacuum solutions obtained in this
paper does not depend on restrictions on parameter $a$. In contrast to these
vacuum solutions the vacuum solution in Ref.  is specific solution. In the
frame of isotropic cosmology without higher derivatives ($a=0$) based on
gravitational Lagrangian (1) (see Refs. 7-15 and references herein) similar
vacuum solutions do not appear. Because we have two possibilities for the
vacuum -- Minkowski spacetime and de Sitter spacetime with torsion, the
following question appears: which of these possibilities is realized in nature
(by assuming that the PGTG is correct gravitation theory). The answer to this
question depends on the behaviour of cosmological solutions for HIM at
asymptotics, when energy density of gravitating matter tends to zero.
Mathematically the answer to the formulated question depends on restrictions on
indefinite parameters in gravitational equations for HIM.

Now we will obtain the vacuum solution at some physically acceptable
restrictions on indefinite parameters. In the frame of isotropic cosmology
without higher derivatives, the absence of which is ensured by condition $a=0$,
equations for HIM include in general case four indefinite parameters (see for
example Ref. 13): $\alpha=\frac {f} {3f_0^2}$ with inverse dimension of energy
density ($f>0$), $b$ with the same dimension as $f_0$ and two dimensionless
parameters: $\varepsilon =\frac{q_2}{f}$ and $\omega=\frac{2f -q_1-q_2}{f}$
\footnote{The parameter $\omega$ was supposed to be equal to zero in Refs.
12-14.}. By some restrictions on indefinite parameters the cosmological
equations take at asymptpotics the form of Friedmann cosmological equations of
GR with effective cosmological constant induced by spacetime torsion and allow
to explain the acceleration of cosmological expansion at present epoch without
using notions of dark energy and also dark matter (at least partially)
\cite{a13}. The behaviour of cosmological solutions depends essentially on
restrictions on parameters $\varepsilon$ and $\omega$. The most simple from
mathematical point of view and physically acceptable case corresponds to the
following choice: $\varepsilon=0$ and $\omega \neq 0$. In this case the
equations (12) and (10) lead to the following vacuum solutions for $S_2^2$ and
$H^2$: $S_2^2= \left [1 - \frac{b}{2f_0}[1 \pm \left(1+ \omega (1-b/f_0)
\frac{f_0^2}{b^2} \right)^{1/2}]\right ][12\alpha b(1- \omega/4)]^{-1}$; $H^2=
\frac{6b^2} {f_0} \alpha S_2^4 [1 - 6 \alpha (2b -\omega f_0) S_2^2]^{-1}$. The
answer to the question "Which of these solutions corresponds to the true
vacuum?" depends on additional restrictions on indefinite parameters and also
on properties of equation of state of gravitating matter at the beginning of
cosmological expansion, by which the cosmological equations lead to regular
cosmological solutions with corresponding asymptotics \cite{a27}.

\section{Conclusion}

In the framework of the standard gravitation theory (GR) de Sitter spacetime
appears as a result of introducing of cosmological constant, which corresponds
to some gravitating object with negative pressure, into Einstein gravitation
equations \footnote{In the framework of PGTG de Sitter spacetime induced by
cosmological constant was considered in Ref. 19.}. By usual interpretation, the
cosmological constant is associated with the vacuum of quantized matter fields.
It should be noted that in the frame of quantum field theory the vacuum energy
density of quantized fields diverges, and it can be eliminated by means of
regularizing procedure. At the same time the value of cosmological constant in
standard $\Lambda CDM$-model corresponds to very small energy density
comparable with average energy density in the Universe at present epoch.

As it was shown in this paper, in the framework of PGTG de Sitter
spacetime appears as a result of exact solution of gravitational
equations for HIM with two torsion functions without cosmological
constant. If the spacetime in the vacuum has the structure of de
Sitter spacetime with torsion, then in the framework of classical
field theory the conception of the vacuum as physical notion is
changed essentially. Instead of vacuum as passive receptacle of
physical objects and processes, the vacuum assumes a dynamical
properties as a gravitating object. That leads to principal
differences of gravitational interaction in comparison with other
fundamental physical interactions, which are connected with
certain matter properties and manifestation of which disappears
without physical matter. In contrast to this, the vacuum possesses
important characteristics of gravitating objects -- the curvature
and torsion.

If the spacetime of the vacuum is de Sitter spacetime with torsion, the
acceleration of cosmological expansion at present epoch explained in the frame
of PGTG in Refs. 12 and 15 acquires the vacuum origin. As it was noted in Ref.
13, the search for the criteria that allow us to be able to choose physically
acceptable solutions is important for PGTG. According to Ref. 5 any vacuum
solution of Einstein gravitation equations of GR (in particular, the
Schwarzschild vacuum solution) with vanishing torsion is an exact solution of
PGTG independently on values of indefinite parameters of gravitational
Lagrangian (1). If the physical spacetime in the vacuum possesses the torsion,
such solutions are not physically acceptable, and the search of corresponding
physically acceptable solutions becomes well warranted. Because in the frame of
PGTG Newton's law of gravitational interaction can be not applicable at
cosmological and possibly astrophysical scales, approximative analysis of
solutions of PGTG, in the frame of which Newton's law is used in the lowest
approximation, has to be re-examined. It should be noted that the analysis of
the particle content of PGTG based on general expression of gravitational
Lagrangian in Refs.5 and 21-23 is given in torsionless backgrounds -- Minkowski
spacetime and Einstein manifolds. If the physical vacuum has the structure of
de Sitter spacetime with torsion, the investigation of the particle content of
PGTG in such background is of certain interest.

We see that the Poincar\'e gauge theory of gravity leads to principal
consequences concerning the classical notion of physical vacuum. By certain
restrictions on indefinite parameters of gravitational Lagrangian (1), unlike
some other generalizations of Einsteinian gravitation theory (see for example
Refs. 25 and 26) PGTG is free of such pathological objects as ghosts and
tachyons, and cosmological solutions for accelerating Universe are
asymptotically stable.

\end{document}